\documentclass[twocolumn,twoside]{svmultivs_gm} 
\usepackage{graphicx}

\usepackage{color}
\definecolor{Dred}{rgb}{0.312,0.070,0.070}
\definecolor{Dblue}{rgb}{0.070,0.070,0.312}
\definecolor{Dgreen}{rgb}{0.070,0.312,0.070}
\definecolor{Db}{rgb}    {0.050,0.0,0.320}

\title*{Progress on VLBI Ecliptic Plane Survey}
\titlerunning{Progress on VEPS}
\author{Fengchun Shu$^1$, Leonid Petrov$^2$, Wu Jiang$^1$, Jamie McCallum$^3$, Sang-oh Yi$^4$, Kazuhiro Takefuji$^5$, Jinling Li$^1$, Jim Lovell$^3$}
\authorrunning{Shu et al.}
\authoremails{sfc@shao.ac.cn, Leonid.Petrov@lpetrov.net, jiangwu@shao.ac.cn, Jamie.McCallum@utas.edu.au, sangoh.yi@gmail.com, takefuji@nict.go.jp, jll@shao.ac.cn, jim.lovell@utas.edu.au}
\institute{1. Shanghai Astronomical Observatory, Chinese Academy of Sciences, China \\
           2. Astrogeo Center, USA \\
           3. University of Tasmania, Australia \\
           4. National Geographic Information Institute, South Korea \\
           5. National Institute of Information and Communication Technology, Japan}
\ContactAuthorName{Fengchun Shu}
\ContactAuthorTelephone{+86 21 34775506}
\ContactAuthorEmail{sfc@shao.ac.cn}
\NumberofInstitutions{1}
\InstitutionPostAddress{1}{80 Nandan Road, Shanghai 200030}
\InstitutionCountry{1}{China}
\InstitutionWebPage{1}{http://www.shao.ac.cn}
\InstitutionPostAddress{2}{N.A}
\InstitutionCountry{2}{N.A}
\InstitutionWebPage{2}{N.A}
\begin{document}  
\maketitle       

\abstract{We launched the VLBI Ecliptic Plane Survey program in 2015. The goal of this program is to find all compact sources within 7.5 degrees of the ecliptic plane which are suitable as phase calibrators for anticipated phase referencing observations of spacecrafts. We planned to observe a complete sample of the sources brighter than 50 mJy at 5 GHz listed in the PMN and GB6 catalogues that have not yet been observed with VLBI. By April 2016, eight 24-hour sessions have been performed and processed. Among 2227 observed sources, 435 sources were detected in three or more observations. We have also run three 8-hour segments with VLBA for improving positions of 71 ecliptic sources.}
%
%
\section{Introduction}
This paper presents the status report of the ongoing VLBI Ecliptic Plane Survey (VEPS) program.
The first goal of the program is to search for
more ecliptic calibrators with a minimum network of 3
stations. We consider a source that is brighter than 30 mJy at baseline projection length
5000 km as a calibrator. We have selected all objects within $7.5^\circ$ of the ecliptic plane, with single dish flux densities brighter than 50~mJy at 5~GHz from the PMN (Parkes-MIT-NRAO) and GB6 (Green Bank 6~cm) catalogues except for those:

\begin{itemize}
\item that have been detected with VLBI before
\item that were observed with VLBI in a high sensitivity mode (detection limit better than 20mJy), but have not been detected.
\end{itemize}

As the target sources number is more than 7000, we planned to observe in two phases.

\begin{itemize}
\item Phase-A --- Observations of 2216 sources that have total flux density at 5~GHz $>$ 100~mJy
\item Phase-B --- Observations of 4802 sources that have total flux
  density at 5~GHz in the range [50, 100]~mJy
\end{itemize}

More details about observations of this large sample of target sources will be described in section 2-4.

The second goal of the VEPS program is to improve the position accuracy better than
1.5 nrad for those ecliptic calibrators detected in various VLBI
experiments, but with large position uncertainties. This type of observations should be performed at S/X dual
band and use a large network such as VLBA, EVN or IVS in a high sensitivity mode. We will address this issue in section 5.
\section{Observations}
The phase-A observations began in February 2015. The participating
stations include the 3 core Chinese VLBI stations: Seshan25, Kunming
and Urumqi. However, sometimes they cannot be available at the same time,
or occasionally one or two of them have a risk of failure. In that
case, one or two international stations are required.

\begin{figure}[htb!]
  \includegraphics[width=.5\textwidth]{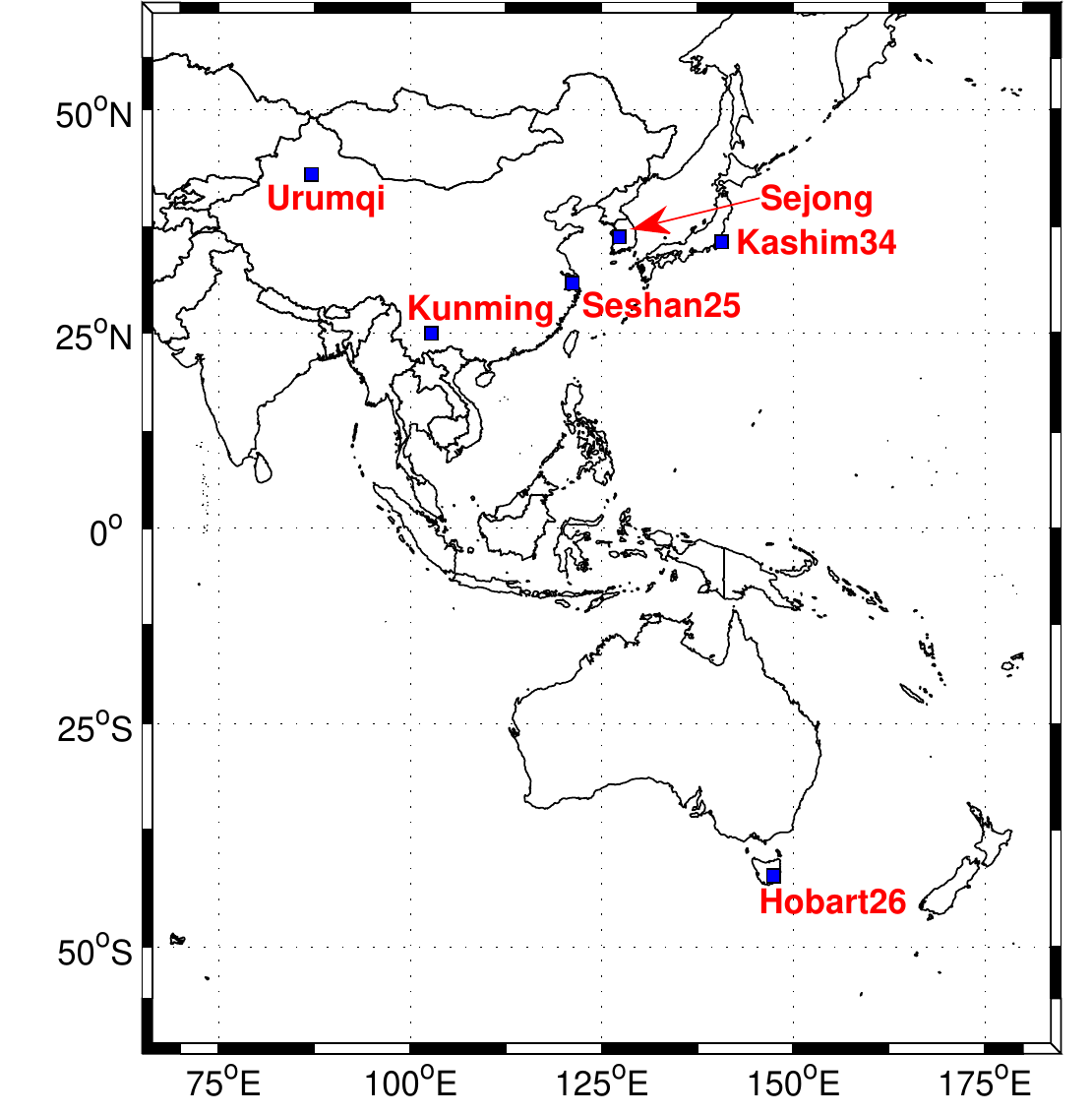}
  \caption{Distribution of participating stations.}
  \label{map}
\end{figure}

Fig.\ref{map} shows the geographical distribution of all
participating stations. Kashima34, Sejong and Hobart26 have
contributed to the past VEPS observations. They have middle-sized
antennas, and good common visibility for the ecliptic zone. Before
joining in the VEPS survey, we made fringe tests to
Sejong, Hobart26 and Kashim34 in December 2014, July 2015, and January
2016 respectively.

So far 8 sessions have been observed, as summarized in
Table~\ref{sessions}. Each target source was observed in two scans of
90 seconds. 4 calibrators were observed every 1 hour for reduction of atmospheric
effects and amplitude calibration.

\begin{table*}[htb!]
\caption{Summary of the VEPS observations}
\begin{tabular}{c|c|c|c|l|c|c|c|c|c|c}
\hline
Date         & Time     & Dur & Code   & Stations & Frequency & Data rate & Channels & Sampling   & Data volume & \# Targets \\
(yyyy-mm-dd) & UT       & hrs &        &          &           & (Mbps)    &          & (bits)     & (TB)   &          \\
\hline
2015-02-13   & 05h00m   & 24 & VEPS01 & ShKmUr	  & X  & 2048 & 16  & 2 & 48      & 293   \\
2015-02-14   & 06h00m   & 24 & VEPS02 & ShKmUr    & X  & 2048 & 16  & 2 & 48      & 338   \\
2015-04-23   & 05h00m   & 24 & VEPS03 & ShKmUrKv  & X  & 2048 & 16  & 2 & 56      & 300   \\
2015-04-24   & 06h00m   & 24 & VEPS04 & ShKmUrKv  & X  & 2048 & 16  & 2 & 56      & 400   \\
2015-08-10   & 05h00m   & 25 & VEPS05 & ShKmKvHo  & X  & 2048 & 16  & 2 & 42      & 252   \\
2015-08-19   & 05h00m   & 25 & VEPS06 & ShKmKvHo  & X  & 2048 & 16  & 2 & 42      & 277   \\
2016-03-02   & 08h30m   & 24 & VEPS07 & ShKmUrKb  & X  & 2048 & 16  & 2 & 52      & 333   \\
2016-03-11   & 05h00m   & 24 & VEPS08 & ShKmUrKb  & X  & 2048 & 16  & 2 & 60      & 477   \\
\hline
\end{tabular}
\label{sessions}

\footnotesize{
Note 1. --- Sh: Seshan25; Km: Kunming; Ur: Urumqi; Kv: Sejong; Kb: Kashim34; Ho: Hobart26.

Note 2. --- The mode 1024 (data rate) - 16 (channels) - 1 (bit) was used at Sejong.

Note 3. --- The mode 1024 (data rate) - 16 (channels) - 2 (bit) was used at Hobart26.
}
\end{table*}

\begin{figure}[htb!]
   \includegraphics[width=.5\textwidth]{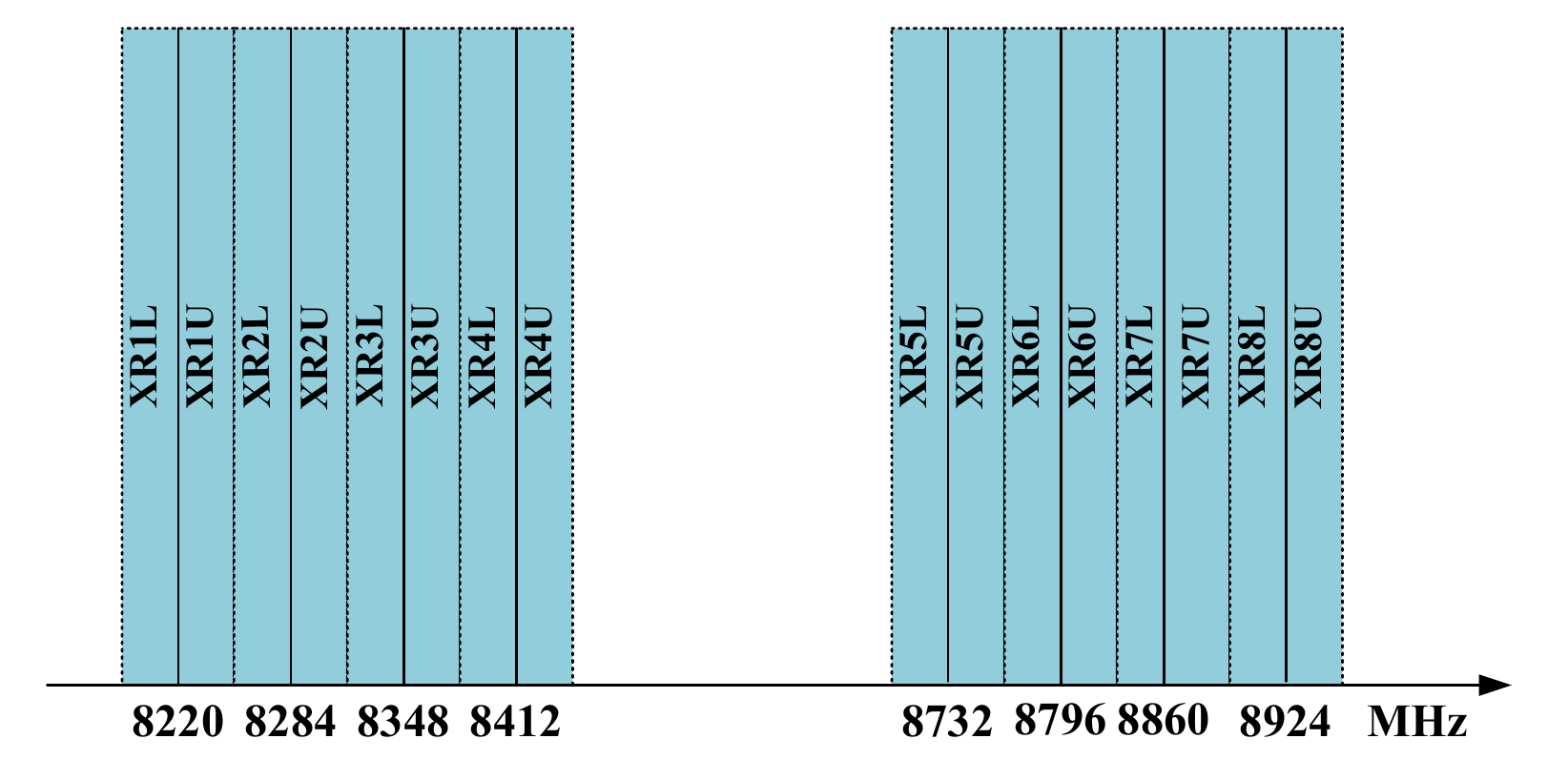}
  \caption{Frequency sequence used in the VEPS observations}
  \label{freq}
\end{figure}

Fig.\ref{freq} shows the frequency sequence used in the phase-A
observations. 8 USB channels and 8 LSB channels spread over about 800
MHz frequency range at X-band result in 16 IF channels, and the
bandwidth for each IF channel is 32 MHz, so the total data rate is
2048~Mbps with 2 bit sampling. The data volume is close to 16 TB for
each station in one 24~h session.

For Sejong station, the maximum rate is 1024 Mbps, so the data
sampling was changed to 1 bit. For Hobart26, its 32~MHz bandwidth had
not been tested at that time, so we observed the first 16~MHz
bandwidth for each IF channel instead. In the case of 1024~Mbps data rate,
the data volume is close to 8 TB for each station in one session.

\section{Data processing}
The data from the Chinese domestic stations were recorded on 16-TB
diskpacks and then shipped to Shanghai, while the data from
international stations were transferred to Shanghai via high speed
network.  The data volume for each session is much bigger than that of
regular geodetic sessions, so the data processing is very
time-consuming. Another technical issue is the correlation of mixed
observing modes with different bandwidths or sampling bits. This could
be supported by the DiFX correlator installed at Shanghai, which
also serves as one of the IVS correlators.

For the correlation of 1-bit sampled data from Sejong against 2-bit
sampled data from the other stations, a different treatment was
implemented in the station based processing module, and it turned to
be the same after the data being transformed to the frequency domain.

For the correlation of 16~MHz bandwidth data from Hobart26 against 32~MHz
bandwidth from the other stations, the zoom mode was selected to pick up
the overlapped frequency band. Morever, it was optional to make
correlation only on the 16~MHz bandwidth on the baselines to Hobart26,
while the other stations with 32~MHz bandwidth went through an
independent correlation pass, the same as the usual correlation
procedures.
\section{Preliminary results}
In general, the VEPS observing sessions were successful, though Urumqi
had no fringes in the first half of VEPS01 session due to a receiver
problem, and Seshan25 and Kunming had no fringes in VEPS03 due to
incorrect use of B1950 source positions.

We have processed all of the observed sessions. Based on data analysis, there
are 435 target sources that were detected in three or more
observations among 2227 observed. The detection rate is about
20\%. Their median position precision is about 18~nrad. The estimation
of the correlated flux densities is better than 15\%.

Except for the baselines to Sejong, the other baselines have detection
limits better than 30~mJy. Deduced from the 4 VEPS sessions, the SEFD
of Sejong varied from 3000 to 5000, which can also be confirmed by the
IVS sessions it participated in. The causes are under investigation
and may be related to the antenna pointing model, aperture efficiency
as a function of frequency or the receiver system.

\begin{figure*}[htb!]
   \includegraphics[width=16.0cm]{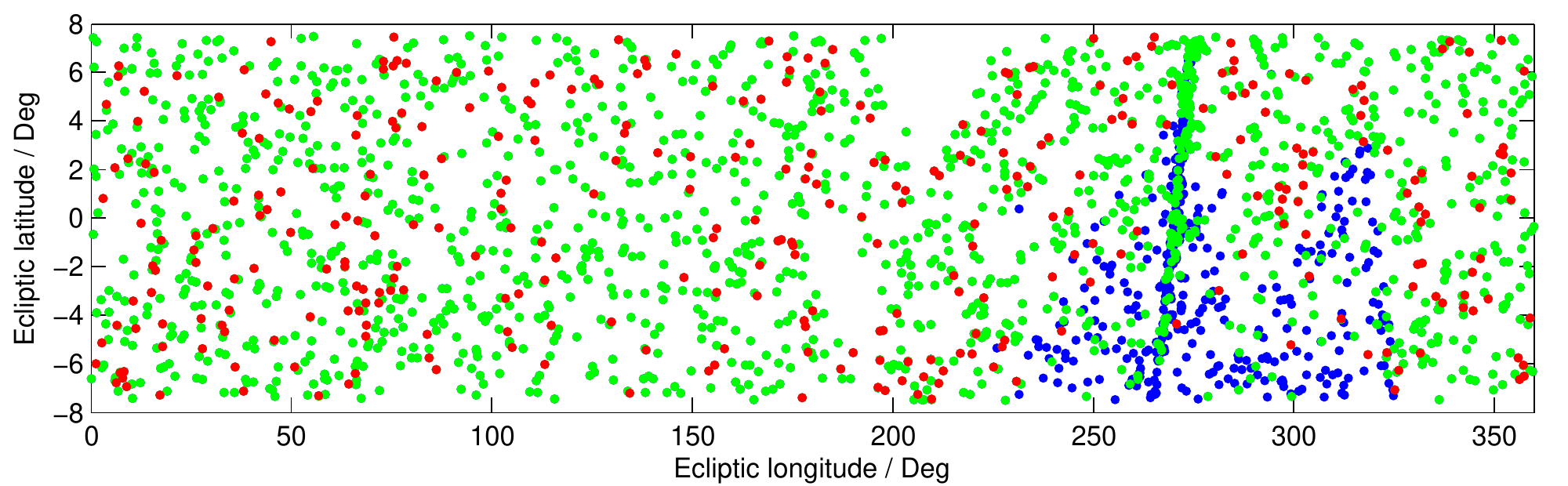}
  \caption{Distribution of Phase-A sources. Among 2216 target sources (\textcolor{blue}{blue}), 1903 sources (\textcolor{green}{green}) have been observed, 363 sources (\textcolor{red}{red}) were detected in 3 or more observations.}
  \label{veps100}
\end{figure*}

\begin{figure*}[htb!]
  \includegraphics[width=16.0cm]{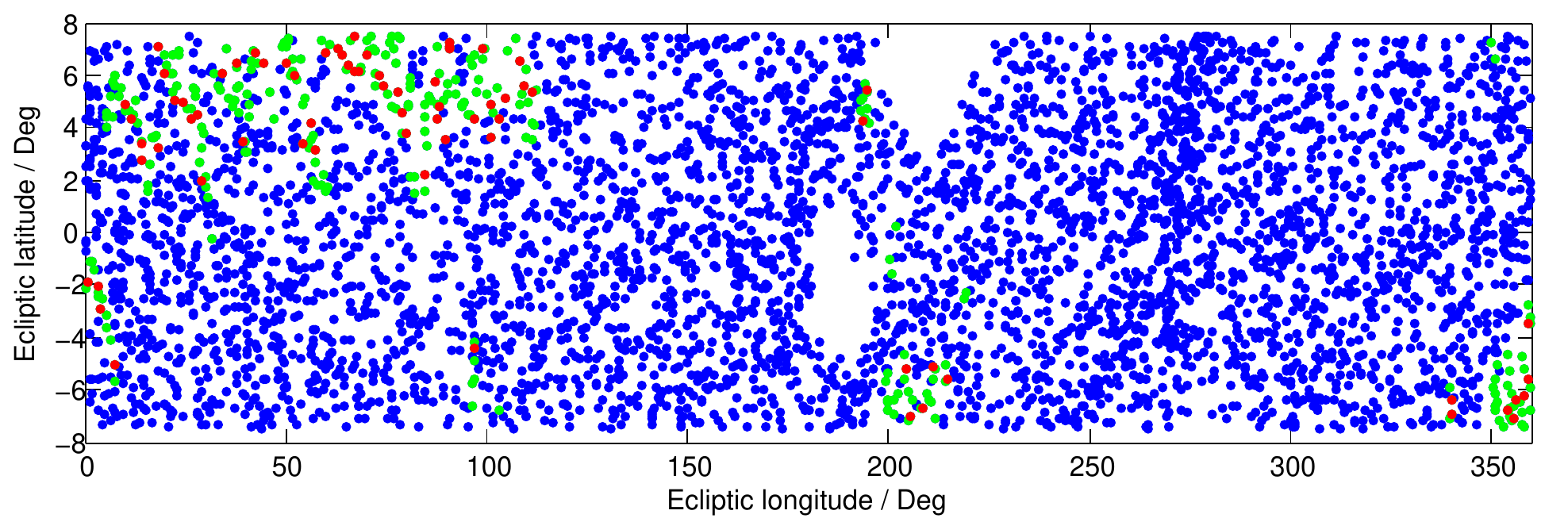}
  \caption{Distribution of Phase-B sources. Among 4802 target sources (\textcolor{blue}{blue}), 324 sources (\textcolor{green}{green}) have been observed, 72 sources (\textcolor{red}{red}) were detected in 3 or more observations.}
  \label{veps50}
\end{figure*}

Fig.\ref{veps100} and Fig.\ref{veps50} shows the distribution of
Phase-A sources and Phase-B sources respectively. Most Phase-A
sources have now been observed. We can see there are two holes in the
plots beside 200 degrees of ecliptic longitude. In the PMN surveys,
these small regions were severely affected by solar contamination when
the sidelobes of the antenna were encountering the Sun, so those data
have been expunged from the survey \cite{pmn}. In the next VEPS
sessions, we will try to fill the two holes with sources from other
radio catalogues.


In order to finish the survey of remaining sources, an additional
400-hours of observing time will be required. We expect that Sejong
will have a better performance with improved SEFD and Hobart26 will
use the DBBC2 2~Gbps mode.
\section{High sensitivity astrometry}
\begin{figure}[htb!]
   \includegraphics[width=.5\textwidth]{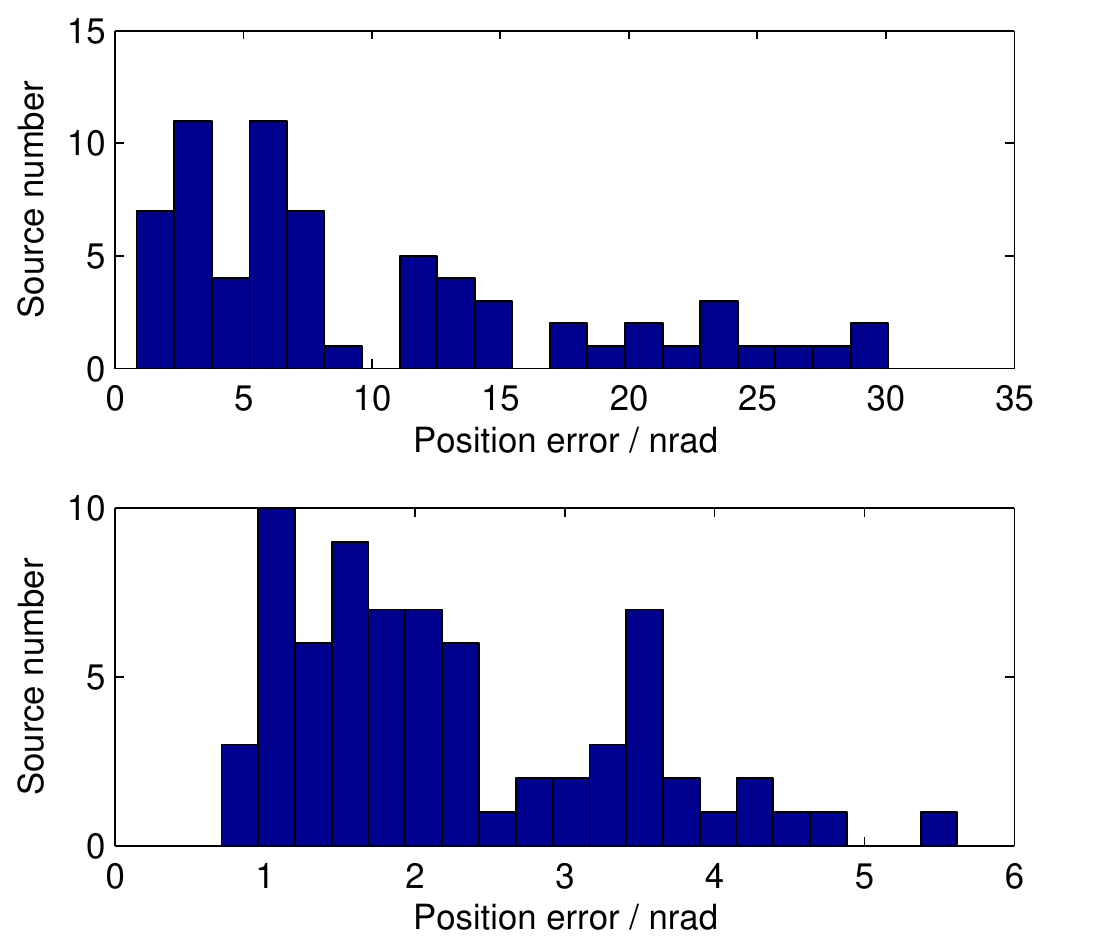}
  \caption{Top: Histogram of position errors of 71 sources before the VLBA observations. Please note 4 sources with position error larger than 35~nrad are not included in the statistics. Bottom: Histogram of position errors of 71 sources after the VLBA observations.}
  \label{bs250}
\end{figure}

As of April 2016, we have observed 73 ecliptic sources with VLBA in three 8-hour segments at 2.3 and 8.6 GHz at 2Gbps (project code: BS250). The targets are the weakest calibrators with correlated flux density at baseline project lengths 5000 km in a range [30, 50] mJy. A priori positions of one half the targets were derived from single-band VLBI observations at 4 or 8~GHz. We scheduled each targets in three scans of 180s long. Two targets have not been detected at S-band. Position uncertainties of 71 remaining targets before our VLBA observations were in a range of [0.8, 294]~nrad with median 6.2~nrad. After our VLBA observations, the position uncertainties dropped to the range of [0.7, 5.6]~nrad, with median 1.8 nrad (See Fig.\ref{bs250}).

Statistics of VLBI detected sources within $\pm 7.5^\circ$ of the ecliptic plane are shown in Table~\ref{ecl}. The number of known calibrators in the ecliptic plane is growing rapidly and reached 1167 recently. Positions of only 23\% have been determined with accuracy better than 1.5 nrad using S/X dual-band VLBI. Changes on 2016-04-01 with respect to 2016-02-01 are contributed by some VEPS, VCS9, and VCS-II experiments \cite{dg}. Changes on 2016-05-01 with respect to 2016-04-01 are contributed by the project BS250.

\begin{table*}[htb!]
\caption{Statistics of sources detected with VLBI within $\pm 7.5^\circ$ of the ecliptic plane. }
\begin{tabular}{l|l|l|l|l}
\hline
                                        & 2015-01-01& 2016-02-01 & 2016-04-01  & 2016-05-01 \\
\hline
\# Calibrators with errors $<$ 1.0 nrad   &      130  &       133  &       143   &      155 \\
\# Calibrators with errors $<$ 1.5 nrad   &      187  &       191  &       261   &      293 \\
\# Calibrators with errors $<$ 2.0 nrad   &      279  &       283  &       405   &      449 \\
\# Calibrators with errors $<$ 2.5 nrad   &      336  &       349  &       479   &      533 \\
\# Calibrators with errors $<$ 3.0 nrad   &      402  &       423  &       518   &      574 \\
\# Calibrators with errors $<$ 5.0 nrad   &      521  &       549  &       625   &      681 \\
Total \# calibrators                      &      772  &       969  &      1167   &     1167 \\
Total \# sources including non-calibrators &    1154  &      1450  &      1710   &     1732 \\
\hline
\end{tabular}
\label{ecl}
\end{table*}
In order to improve positions of known calibrators, we plan to observe them with VLBA, EVN, or IVS.
In a more practical sense, it might be a good
idea to form a high sensitivity network with Asian and Oceanian
antennas. In such a network, a compatible observing mode using S/X
band at 2~Gbps needs to be defined and tested. With
inclusion of the Tianma or Parkes radio telescopes, even the baselines
to small antennas of AuScope will have sensitivity comparable with VLBA.
\section{Conclusions}
The VEPS observations for detecting more ecliptic calibrators are running
smoothly and have become routine work now. More than 400 sources have been
detected on VLBI baselines for the first time.

The VEPS observations for improving positions of known calibrators will benefit
from a proposed high sensitive network with Asian and Oceanian
antennas.
%


\section*{Acknowledgements}
The work is carried out with the support of the National Natural Science Foundation of China (U1331205).

%
%
%

%
\end{document}